\documentclass[a4paper]{article}

\usepackage{icrc2013}
\usepackage[english]{babel}

\title{VER J2019+407 and the Cygnus Cocoon}

\shorttitle{VER J2019+407 and the Cygnus Cocoon}

\authors{
Joshua V. Cardenzana$^{1}$
for the VERITAS Collaboration.
}

\afiliations{
$^1$Iowa State University, Department of Physics \& Astronomy \\
}

\email{jvcard@iastate.edu}

\abstract{Current gamma-ray observatories have made substantial advances in the study of potential cosmic ray accelerators such as supernova remnants (SNRs), but the study of cosmic rays in flight from their parent accelerators remains a topic of great interest. The $Fermi$ Gamma-ray Space Telescope (FGST) discovered a four square degree region of gamma-ray emission between 3 GeV and 100 GeV. This structure lies between the Cygnus OB2 association and the $\gamma$-Cygni SNR, which is bright in gamma rays above 10 GeV. The atmospheric Cherenkov observatory VERITAS also sees VHE ($>$100 GeV) gamma-ray emission from the direction $\gamma$-Cygni (VER J2019+407). We discuss the relationship of VER J2019+407 to both the $\gamma$-Cygni SNR and the cocoon. Toy model simulations are used to evaluate prospects for detecting the cocoon at very high energies with VERITAS, as well as ways in which cocoon emission at these energies could impact measurements of the morphology and spectrum of VER J2019+407.}


\begin{document}
\maketitle


\section{Introduction}

The cosmic rays that pervade our galaxy are thought to have escaped from astrophysical accelerators that accelerate them, such as supernova remnants (SNRs).  Direct observations of cosmic rays provide little information as to their origin, since the paths of cosmic rays are distorted by interstellar magnetic fields.  However, cosmic-ray accelerators may be mapped and studied indirectly through high energy (300 MeV - 100 GeV) and very high energy (100 GeV - 30 TeV) gamma-ray observations.  Particles still trapped within these accelerators are expected to produce gamma-ray emission through a number of mechanisms.  Accelerated electrons may produce gamma-ray emission through synchroton radiation and inverse Compton (IC), while interactions of accelerated protons and heavier nuclei with the interstellar medium may produce gamma rays through neutral pion decay.  Cosmic rays recently escaped from their parent accelerators may also produce recognizable signatures of diffuse gamma-ray emission.

\section{$Fermi$-LAT Observations of the Cygnus Region}

Recent observations with the Large Area Telescope (LAT) on the $Fermi$ Gamma-ray Space Telescope have revealed a region of highly extended gamma-ray emission \cite{bib:Ackermann}, best characterized by a Gaussian with $\sigma$=2.0$^{\circ}$ $\pm$ 0.2. Flanked on the north by the supernova remnant SNR G78.2+2.1 ($\gamma$-Cygni) and on the south by the massive star cluster Cygnus OB2, the emission is believed to be a cocoon of freshly accelerated cosmic rays. While the actual origin of the cosmic rays within the cocoon is unknown, it is possible they were originally produced in either or both of these two sources.

$Fermi$ also sees diffuse emission above 10 GeV from the entire $\gamma$-Cygni SNR \cite{bib:Lande}, as well as a gamma-ray pulsar, PSR J2021+4026, at the center of the remnant \cite{bib:Trepl}. On the other side of the cocoon $Fermi$ observes a pulsar coincident with TeV J2032+4130 in the Cygnus OB2 region \cite{bib:Abdo2010}. Study of this region in X-ray and gamma-ray regimes could therefore provide valuable insight into the nature of recently escaped cosmic rays.

\section{The VERITAS Instrument}

VERITAS consists of four 12m imaging atmospheric Cherenkov telescopes which achieved first light in 2007. The instrument is sensitive within the energy range of 85 GeV to 30 TeV. The array is able to achieve single photon angular resolution of 0.1$^{\circ}$ (68\% containment) and an energy resolution of 15-25\%. Observations cover a 3.5$^{\circ}$ field of view. For further information see Holder et al.\cite{bib:Holder}.

\section{Cocoon Region as seen by VERITAS}

\begin{figure}[t]
  \centering
  \includegraphics[width=0.5\textwidth]{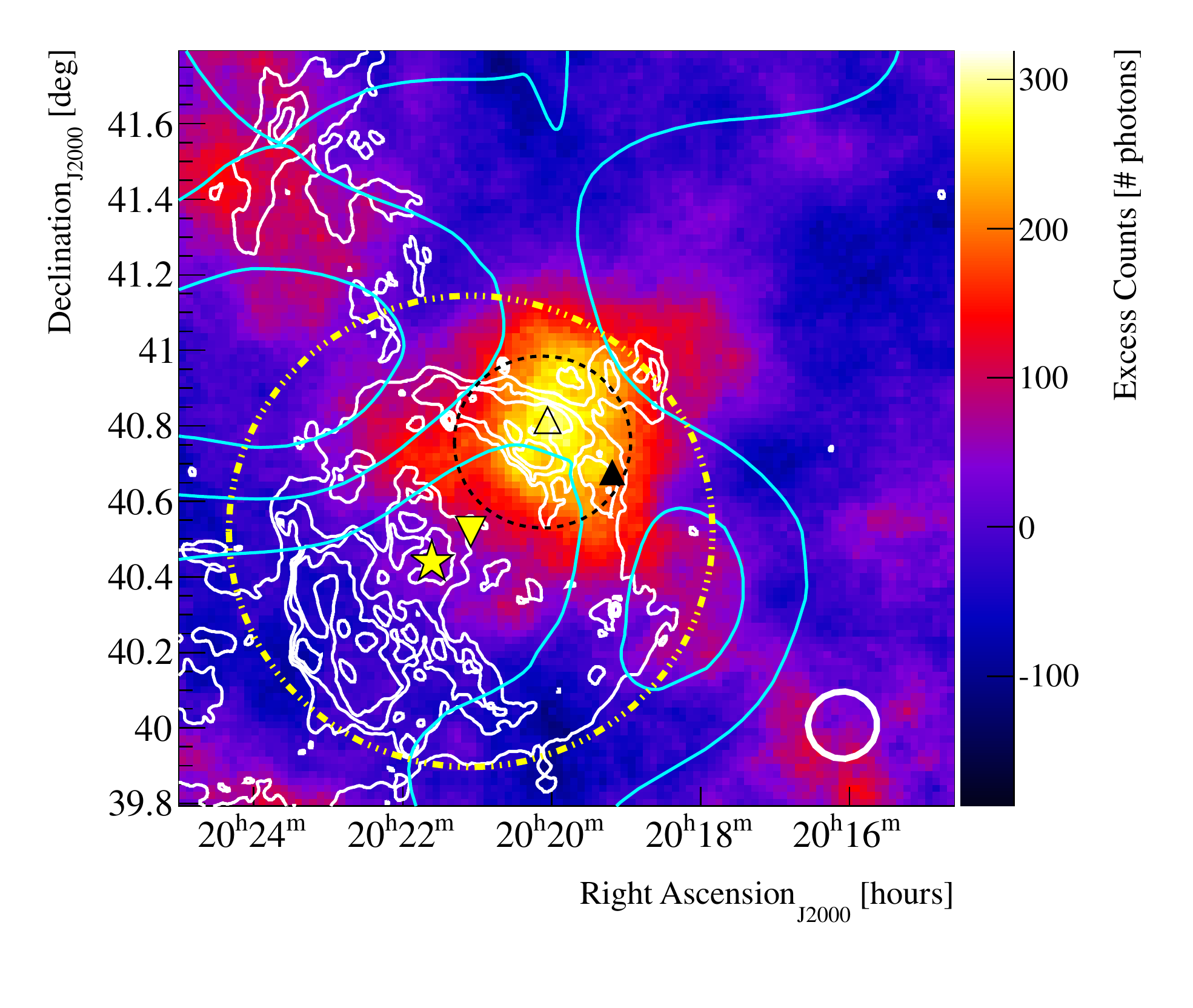}
  \caption{Background subtracted $\gamma$-Cygni region showing the bright emission from VER J2019+407 and its extent (black dashed circle). $Fermi$-LAT results plotted are the fitted extent of $\gamma$-Cygni as seen above 10 GeV (yellow) and the cocoon at 0.16, 0.24, and 0.32 photons bin$^{-1}$ (cyan). The 68\% containment size of the VERITAS gamma-ray PSF is also shown (lower-right corner). Image reproduced from Aliu et al. (2013)\cite{bib:Aliu}.}
  \label{g-Cygni}
 \end{figure}

VERITAS sees two gamma-ray sources in the vicinity of the cocoon \cite{bib:Weinstein09}, both related to already-noted potential cosmic ray accelerators. The first is TeV J2032+4130, a previously known source of VHE gamma-rays first observed with HEGRA \cite{bib:Aharonian} and associated with Cygnus OB2. The second is VER J2019+407, a previously unknown extended source co-located with the $\gamma$-Cygni SNR \cite{bib:Aliu} (figure \ref{g-Cygni}).

Indications of VER J2019+407 were first seen by VERITAS in a survey of the Cygnus region of the Galactic plane, conducted in 2008-2009.  The source was discovered in follow-up observations taken in 2009. The source is notably extended, being well-fit by a symmetric Gaussian with $\sigma$ = 0.23$^{\circ}$ $\pm$ 0.03$_{stat}$ $_{-0.02}^{+0.04}$$_{sys}$, convolved with the VERITAS point spread function (PSF). The fitted source spectrum follows a differential power law
\begin{equation}
\frac{dN}{dE} = N_{0} \times \left( \frac{E}{TeV} \right) ^{- \Gamma }. 
\end{equation}
with flux normalization N$_{0}$ = 1.5 $\pm$ 0.2$_{stat}$ $\pm$ 0.4$_{sys}$ $\times$10$^{-12}$ TeV$^{-1}$cm$^{-2}$s$^{-1}$ and photon index $\Gamma$ = 2.37 $\pm$ 0.14$_{stat}$ $\pm$ 0.20$_{sys}$.

Aliu et al. (2013) consider the possibility that the gamma-ray emission from VER J2019+407 is produced by either hadronic interactions of accelerated nuclei with ambient gas or inverse Compton (IC) scattering of ambient light by accelerated electrons. They calculate that a density of 1.0-5.5 cm$^{-3}$ is required to produce the gamma-ray flux above 320 GeV if the emission is hadronic in nature and note this is consistent with estimates of nearby densities obtained from optical observations. Due to the lack of non-thermal X-ray emission near VER J2019+407, Aliu et al (2013) do not favor an IC origin to the source, but cannot rule out leptonic emission at this time.

VERITAS has not detected evidence of the cocoon itself. This does not necessarily imply that gamma-ray emission above 100 GeV is not present since the methods of background estimation used up to this point are not suited to detection of such largely extended sources.

To demonstrate this we can consider one method for background estimation known as the Ring Background method where the background is measured by integrating the number of counts within an annulus centered at the test location. The normalization of the integrated background is corrected for the radial acceptance of the camera \cite{bib:Aharonian2005}. Consider two observations of the cocoon, one centered on the source, and one offset. In the former scenario it is easy to see that for most choices of annulus size, the estimated background counts will be nearly identical to the counts in the source region near the center of the field of view, since the background region still overlaps the expected emission region. 

The later scenario would result in a less gamma-ray-contaminated field of view allowing for the choice of more sensible background regions, but the overall exposure to the cocoon is subsequently reduced. This is similar to the above observations of VER J2019+407 by Aliu et al. (2013) which overlap with the cocoon region. Since they do not not include the cocoon when modeling this region it is possible that gamma-ray leakage from the cocoon could impact the background estimation for its detection and spectrum. 

\section{VERITAS Prospects in the Cocoon Region}

We propose an alternative method for analyzing VERITAS data in the cocoon region, a maximum likelihood method similar to that used for $Fermi$ source detection. However, in addition to modeling contributions of compact and diffuse gamma-ray sources, a VERITAS likelihood would also require modeling of the irreducible isotropic cosmic ray background. The fit is unbinned in three dimensions---two spatial coordinates and a gamma-hadron discrimination parameter known as mean-scaled-width (MSW) --- and is performed simultaneously in three coarse bins in energy (100 - 500 GeV, 500 GeV - 1 TeV, and $>$1 TeV). These specific energy bins were chosen to accommodate differences in the telescope response at threshold, moderate, and high energies.

Maximum likelihood estimation would then be used to evaluate the significance of a source at a grid of reasonably chosen locations in the field of view to produce a significance map. This is done by calculating a likelihood value for a purely cosmic ray background model (L$_{null}$) and for a model which includes our gamma-ray source model as well as a purely background model (L$_{test}$). These two likelihood values represent the likelihood our data originates from the associated model. A test statistic (TS) can then be computed via

\begin{equation}
TS = -2 \cdot log\left( \frac{L_{null}}{L_{test}} \right).
\end{equation}

We also generate toy Monte Carlo simulations representing real data which can be used to evaluate how this TS value relates to true source significance. This is covered more in section 5.3.

In the case of point sources or slightly extended gamma-ray sources the spatial models for gamma-ray and cosmic ray emission are quite distinct from each other (since cosmic rays are distributed across the whole field of view, only modulated by variations in sensitivity over the camera). In this case the spatial dimensions are sufficient to allow for a good source detection. For highly extended sources ($\sigma$ $\ge$ 1.0$^{\circ}$) the spatial model for gamma-rays differs only slightly from the cosmic ray spatial model. A maximum likelihood analysis based only on spatial information will therefore be insensitive to sources on this scale. In order to increase the sensitivity to highly extended gamma-ray sources we also fit in mean-scaled-width (MSW).

The models used in the fit can also be used to produce toy Monte Carlo simulations of VERITAS observations in order to test the ability of VERITAS to detect emission from the cocoon region. As mentioned above, the actual fit is perfomed simultaneously in three bins in energy. This allows us to accomodate the fact that key parameters of these models, such as gamma-ray PSF and cosmic ray background distributions, in both space and MSW, are energy-dependent.

Unless otherwise stated, standard cuts have been applied to all data used in the following discussion (mean-scaled-length $<$1.3, size $>$400dc, image centroid distance $<$1.43, number of telescopes used in position reconstruction $\ge$3).

\subsection{Spatial Models}

The VERITAS field of view is sufficiently small at 3.5$^{\circ}$ that any intrinsic fluctuations in the cosmic ray background is neglected for an individual pointing. However, the response of the instrument is not uniform across the field of view showing a higher acceptance near the center of the camera while dropping off towards the edge. To model this radial acceptance curve we make use of data containing no detectable gamma-ray component and sum the number of observed events within multiple annuli around the camera center. Each bin is then corrected for the area of the annulus, the histogram fit with a polynomial, and the resulting curve extrapolated into two dimensions (assuming azimuthal symmetry) to produce a spatial cosmic ray model.

In order to model the instrument response to gamma-rays an uncontaminated sample of gamma-rays was needed. Use was made of simulations assuming a point source observed by the VERITAS detector and processed in the same way as VERITAS data. The resulting distribution was then fit with a Gaussian obtaining a sigma between 0.05-0.09 depending on the energy range of the fit. It should be noted that a more rigorous treatment of the gamma-ray distribution will consider certain detector specific properties such as energy resolution and effective area as outlined by Mattox et al. \cite{bib:Mattox}. A Gaussian is, however, a good approximation for our proof of principle study. For an extended source this gamma-ray point source model is then convolved with the expected source extension model, which for the Cygnus cocoon is a Gaussian with $\sigma$=2.0$^{\circ}$.

To remove edge effects in these models, events with a reconstructed position greater than 1.7$^{\circ}$ from the camera center are not used in the fits. As a result, both cosmic ray and gamma-ray spatial distributions are modeled to an angular distance of 1.7$^{\circ}$ from the center of the field of view.

\subsection{Mean-Scaled-Width Models}

MSW is a parameter which relates the average image width of an event as observed by each VERITAS telescope to the expected image width from simulations \cite{bib:Hillas}. The distribution of MSW is narrowly peaked around 1 for gamma-ray events.

\begin{figure}[t]
  \centering
  \includegraphics[width=0.4\textwidth]{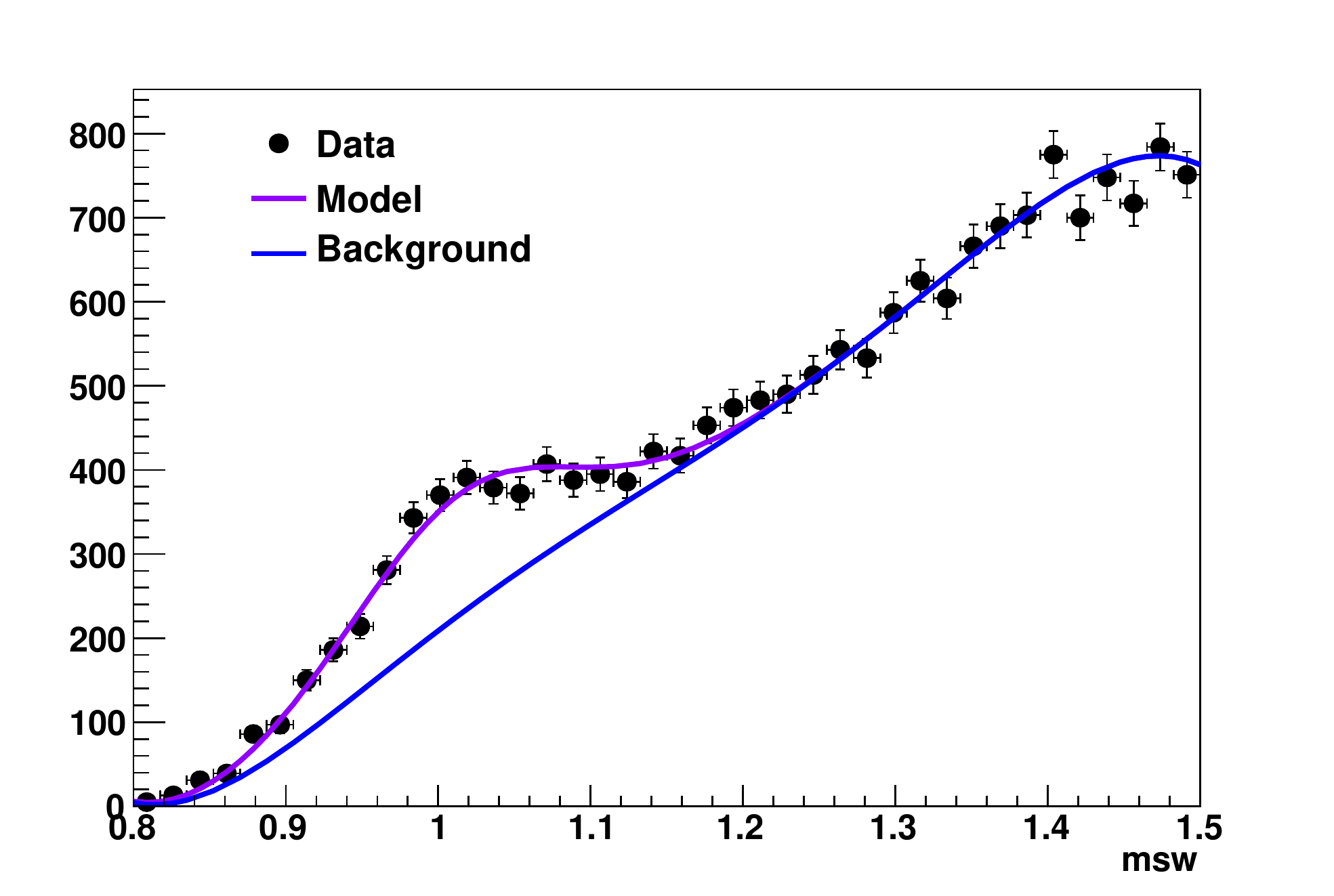}
  \caption{VERITAS data from the Crab Nebula (approximately 10 hours) with a fitted point source model overlay projected in MSW. Both the full model (purple) and the purely background component (blue) are shown.}
  \label{CrabMSW}
 \end{figure}

The distribution for cosmic ray events is much broader and peaks between 1.5-2.0. The cosmic ray distribution is dependent on parameters such as radial offset from the camera center and energy, however for MSW $\le$ 1.5 the shape is sufficiently uncorrelated that we can neglect these effects. As a result we confine our MSW models to the range 0.05-1.5 (see figure \ref{CrabMSW}). It is important to note that while the spatial models and MSW models are correlated at large values of MSW, for the cuts outlined the correlations are small and may be neglected. As a result, we may simply multiply the spatial and MSW pdfs and our full input model in spatial (\textit{x},\textit{y}) and MSW (\textit{w}) is
\begin{eqnarray}
F(x,y,w) = N\cdot[f\cdot S_{src}(x,y)\cdot MSW_{src}(w) \nonumber \\ 
+ (1-f)\cdot S_{bkgd}(x,y)\cdot MSW_{bkgd}(w)]
\end{eqnarray}
where \textit{S$_{src,bkgd}$} represents the spatial models for the source (gamma-rays) or background (cosmic rays) respectively, and \textit{MSW$_{src,bkgd}$} represents the associated MSW models. \textit{N} is the total number of observed events, and \textit{f} is the fraction of the total number of events accounted for by the signal component. These two parameters are determined in our likelihood fits for each energy bin, however in the above expression and the example to follow we only consider the energy range 500 GeV - 1 TeV. This is a simplification relative to the full fit where these two parameters would be replace by spectral parameters such as flux normalization and spectral index and fit across the full energy range of the instrument.

\subsection{Toy Monte Carlo Data Generation}

The resulting cosmic ray and gamma-ray models are combined to form the expected emission model in the cocoon region. Using a Monte Carlo technique and these idealized models, a toy Monte Carlo simulation of an actual VERITAS observation is produced. These simulated observations can then be processed normally through standard or new techniques to investigate the ability of VERITAS to observe the Cygnus cocoon and surrounding region.

\begin{figure}[t]
  \centering
  \includegraphics[width=0.5\textwidth]{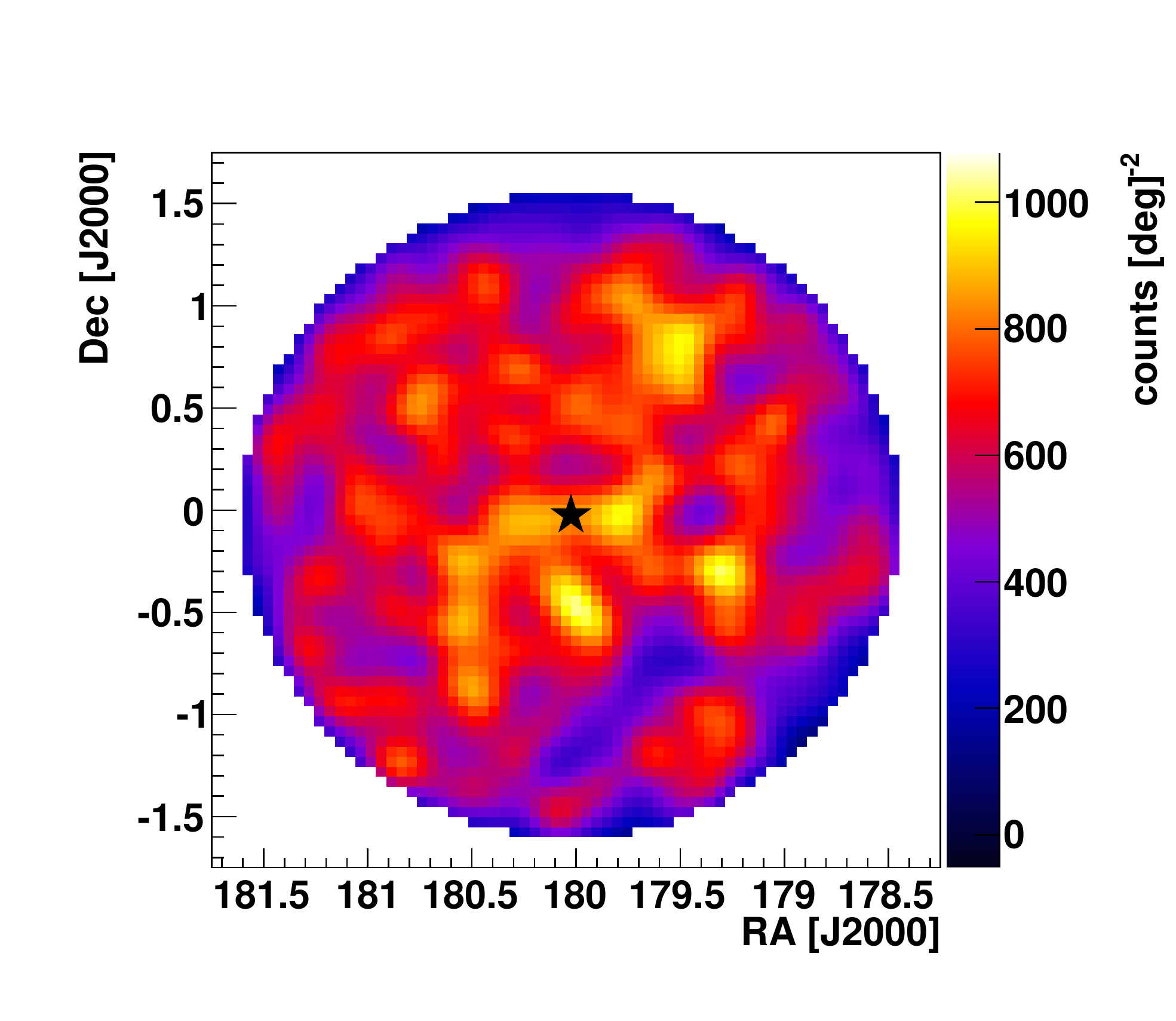}
  \caption{Excess map from toy Monte Carlo simulating a VERITAS observation of the cocoon region integrated over approximately 20 hours.}
  \label{2degExtSrc}
 \end{figure}

As a test of our models, toy Monte Carlo simulating approximately 20 hrs of VERITAS observations of a 2$^{\circ}$ extended emission region similar to the cocoon were generated and fit in a single bin of energy (0.5 - 1 TeV). An estimate on the expected excess signal events was obtained by integrating the cocoon spectrum obtained by Ackermann et al. \cite{bib:Ackermann} in this energy range. The result was then compared to the Crab nebula flux in this energy range and the ratio of the two fluxes multiplied by the best fit \textit{f$_{0.5-1 TeV}$} to real Crab Nebula observations. Using the above MSW and spatial cuts on the toy simulations resulted in an estimated gamma-ray excess of 1.8 min$^{-1}$ across the entire field of view (within a typical RBM source integration window of radius 0.13$^{\circ}$ this would correspond to an excess of 0.011 min$^{-1}$). Fitting only the spatial model (without MSW) to the generated data results in no significant detection. Including the MSW dimension in our fits results in a significance of 27.2$\sigma$. Both models use the same 2$^{\circ}$ Gaussian model that was used for the generation of the data and calculate the significance by taking the square root of the TS value from eqn. 2. The resulting excess map integrated over the full 20 hour simulated observation can be viewed in figure \ref{2degExtSrc}.

\section{Conclusions}

The Cygnus cocoon observed by $Fermi$ presents a rare opportunity to study the origin of Galactic cosmic rays. An intense study of this region could reveal insight as to the nature of cosmic rays still within their parent accelerators, including their composition and energy distributions relative to those still trapped. Understanding the spectrum and morphology of the cocoon at energies above 100 GeV could prove essential to understanding both the population of cosmic rays within the cocoon and the relationship of the cocoon to nearby potential cosmic ray accelerators such as VER J2019+407. While the toy Monte Carlo study discussed here is both idealized and highly preliminary, it indicates that a three-dimensional maximum likelihood method shows promise for detecting sources such as the cocoon that have previously been thought inaccessible (or nearly inaccessible) to small field-of-view  gamma-ray observatories.

\vspace*{0.5cm}
\footnotesize{{\bf Acknowledgment:}{This research is supported by grants from the U.S. Department of Energy Office of Science, the U.S. National Science Foundation and the Smithsonian Institution, by NSERC in Canada, by Science Foundation Ireland (SFI 10/RFP/AST2748) and by STFC in the U.K. We acknowledge the excellent work of the technical support staff at the Fred Lawrence Whipple Observatory and at the collaborating institutions in the construction and operation of the instrument. Dr. Weinstein's and J. Cardenzana's research was supported in part by NASA grant NNX11A086G.}}

\end{document}